\documentclass[preprint,12pt]{elsarticle}

\usepackage{amsthm}
\usepackage{times}
\usepackage{bm}
\usepackage{helvet}
\usepackage{courier}
\usepackage{url}
\usepackage{amsfonts}
\usepackage{amsmath,amssymb}
\usepackage{algorithm}
\usepackage{algorithmic}
\usepackage{color}
\usepackage{xcolor}

\newcommand{\be}{\begin{equation}}
\newcommand{\ee}{\end{equation}}
\newtheorem{theorem}{Theorem}
\newtheorem{lemma}{Lemma}
\newdefinition{claim}{Claim}
\newtheorem{corollary}{Corollary}
\newtheorem{definition}{Definition}
\newproof{prf}{Proof}

\newcommand{\e}{\epsilon}
\newcommand{\bo}{\boldsymbol{o}}
\newcommand{\bp}{\boldsymbol{p}}
\newcommand{\bq}{\boldsymbol{q}}

\DeclareMathOperator*{\argmax}{argmax}

\journal{Computers \& Electrical Engineering}

\begin{document}

\begin{frontmatter}



\title{Fast algorithms for $k$-submodular maximization subject to a matroid constraint}


%
\author[1]{ShuXian Niu}
\ead{niushuxianBetty@163.com}

\author[1]{Qian Liu}
\ead{lq\_qsh@163.com}

\author[1]{Yang Zhou}
\ead{zhouyang@sdnu.edu.cn}

\author[1]{Min Li\corref{cor1}}
\ead{liminemily@sdnu.edu.cn}

\cortext[cor1]{Corresponding author: liminemily@sdnu.edu.cn}

 \address[1]{School of Mathematics and Statistics, Shandong Normal University, 250014, Jinan, China}
%

\begin{abstract}

In this paper, we apply a Threshold-Decreasing Algorithm to maximize $k$-submodular functions under a matroid constraint, which reduces the query complexity of the algorithm compared to the greedy algorithm with little loss in approximation ratio.
We give a $(\frac{1}{2} - \e)$-approximation algorithm for monotone $k$-submodular function maximization, and a $(\frac{1}{3} - \e)$-approximation algorithm for non-monotone case, with complexity $O(\frac{n(k\cdot EO + IO)}{\epsilon} \log \frac{r}{\epsilon})$, 
where $r$ denotes the rank of the matroid, and
$IO, EO$ denote the number of oracles to evaluate whether a subset is an independent set and to compute the function value of $f$, respectively. Since the constraint of total size can be looked as a special matroid, called uniform matroid, then we present the fast algorithm for maximizing $k$-submodular functions subject to a total size constraint as corollaries.
\end{abstract}




\end{frontmatter}


\section{Introduction}

Given a finite set $E = \{ e_{1},e_{2},\ldots,e_{n} \}$, 
let $ 2^{E} $ represent the collection of all subsets of $E$.
Define a set function $ f:2^{E} \to \mathbb{R} $ as a \emph{submodular function} if the following condition is satisfied, 
\[
f(P) + f(Q) \ge f(P \cup Q) + f(P \cap Q),
\]
where $P,Q$ are arbitrary subsets of $E$.
This definition can also be extended to a  $k$-submodular function.

We assume that $k$ is a positive integer, and denote the set $\{ 1,2,\ldots,k \}$ by $[k]$. 
Let $ ( k+1 ) ^E: = \{ ( P_{1},\ldots,P_{k} ) | P_{i}  \subseteq E ( \forall i \in [k] ), P_{i} \cap P_{j} = \emptyset ( i \ne j ) \} $ denote a set of $k$-tuples consisting of $k$ disjoint subsets of $E$.
Similar to the definition of submodular function, we can give the definition of $k$-submodular function as follows:
A set function $ f:(k+1)^E \to \mathbb{R} $ is defined as a \emph{$k$-submodular function} if and only if for any $ \boldsymbol{p} = (P_{1},P_{2},\ldots,P_{k}),  \boldsymbol{q} = (Q_{1},Q_{2},\ldots,Q_{k}) \in (k+1)^{E} $, we have,
\[
f ( \boldsymbol{p} ) + f ( \boldsymbol{q} ) \ge f ( \boldsymbol{p} \sqcup \boldsymbol{q} ) + f ( \boldsymbol{p} \sqcap \boldsymbol{q} ),
\]
where

\begin{align*}
\boldsymbol{p} \sqcap \boldsymbol{q}:&=(P_{1} \cap Q_{1},\ldots,P_{k} \cap Q_{k}),\\
\boldsymbol{p} \sqcup \boldsymbol{q}:&=(P_{1} \cup Q_{1}\backslash(\bigcup_{i \ne 1}P_{i} \cup Q_{i}),\ldots,P_{k} \cup Q_{k}\backslash(\bigcup_{i \ne k}P_{i} \cup Q_{i})).
\end{align*}

Huber and Kolmogorov~\cite{huber2012towards} defined and studied the $k$-submodular polyhedron, then they generalized the Min-Max-Theorem for submodular functions to $k$-submodular functions, which was the first introduction of the concept of $k$-submodular functions.
Ward and \v{Z}ivn\'{y}~\cite{WardZivny2016Beyond} firstly provided the approximation guarantees for maximizing a $k$-submodular function without any constraints. 
When the objective function is monotone, 
they gave a deterministic $\frac{1}{3}$-approximation algorithm based on greedy technique. 
Then Iwata et al.~\cite{IwataTY15} presented a better result by proving the existence of a stochastic $\frac{k}{2k-1}$-approximation algorithm.
Based on this, Oshima~\cite{Oshima2018} gave a de-randomization scheme which yielded the same approximate ratio.  
Wang and Zhou~\cite{Wang2021MultilinearEO} proposed a new technique for analyzing the problem of maximizing a $k$-submodular function, where a multilinear extension of the algorithm that was almost as good as the previous ones was presented. 
When the objective function $f$ is non-monotone, Ward and \v{Z}ivn\'{y}~\cite{WardZivny2016Beyond} gave a stochastic $\max\{\frac{1}{3},\frac{1}{1+\alpha}\}$-approximation algorithm, where $\alpha = \max\{1,\sqrt{\frac{k-1}{4}}\}$. When $k \ge 3$, Oshima~\cite{Oshima2018} obtained a better approximation of $\frac{k}{3k-2}$. 
Iwata et al.~\cite{IwataTY15} improved these results  to a stochastic algorithm with an approximation of $\frac{1}{2}$.  
Later,
Oshima~\cite{HO2019ImprovedRandomized} proposed an improved randomized algorithm based on the ideas of Iwata et al.~\cite{IwataTY15}, that is,
for $k \ge 3$, they gave the algorithm with approximation of  $\frac{k^{2}+1}{2k^{2}+1}$, and for $k = 3$, they presented a stochastic $\frac{\sqrt{17}-3}{2}$-approximation algorithm.

Ohsaka and Yoshida~\cite{Ohsaka2015MonotoneKF} gave approximation algorithms to maximize the monotone $k$-submodular function under two kinds of size constraints: One is total size constraint, the other is individual size constraint.

A total size constraint means that the total number of selected elements in the set of $k$-tuple is limited by a positive integer size $B$, they gave a greedy algorithm with an approximation guarantee of $\frac{1}{2}$ with query complexity of $O(knB)$. 
Combined with random sampling techniques, they also gave a randomized algorithm that can be guaranteed to output an approximation of $\frac{1}{2}$ with probability of at least $1-\delta$, whose query complexity is $O(kn\log B\log\frac{B}{\delta})$.   
Nie et al.~\cite{Nie2023FAST_size} applied the threshold decreasing method combined with greedy ideas to this problem, improving the complexity to $O(\frac{nk}{\e} \log \frac{B}{\e})$ while obtaining a near-optimal approximation, achieving no linear dependence on the budget $B$. 
Nguyen and Thai~\cite{nguyen20f} also proposed streaming algorithms for this kind problems.

Now we introduce the individual size constraint that the number of chosen elements in each of the $k$ sets is limited by a size $B_{i}$, respectively.
For maximizing a monotone subjective function with individual size constraint, Ohsaka and Yoshida~\cite{Ohsaka2015MonotoneKF} gave an algorithm that can output an approximation of $\frac{1}{3}$ in running time $O(knB)$, where $B = \sum_{i=1}^{k}{B_{i}}$. 
Nie et al.~\cite{Nie2023FAST_size} applied the threshold greedy algorithm to this problem, obtaining an approximation ratio of $\frac{1}{3}-\e$ with query complexity of $O(\frac{nk}{\e} \log \frac{B}{\e})$.
They also presented a stochastic $\frac{1}{3}$-approximation algorithm with probability of at least $1-\delta$ with query complexity of  $O(k^{2}n\log{B/k}\log{(B/\delta)})$. 
A new streaming algorithm for this kind of problems was introduced in~\cite{EneAlina2022Streaming}. 
Zheng et al.~\cite{MinmingLi2021} also proved that a simple greedy algorithm to maximize an approximately $k$-submodular function $F$ and obtained reasonable approximation ratios for both kinds of size constraints.
When the objective function $f$ is non-monotone, Xiao et al.~\cite{Xiao2023} gave two kinds of algorithms, one was the $\frac{1}{B_{m}+4}$-approximation algorithm, where $B_{m}=\max\{B_{1},B_{2},\ldots,B_{k}\}$, the other was a bi-criteria algorithm with approximation ratio of $\frac{1}{4}$.
Shi et al.~\cite{FASTshi2021k} proposed a $(1-\frac{1}{e}-\e)$-approximation algorithm with running time of $O(\frac{nk!}{\e}\log \frac{n}{\e})$.

Sakaue~\cite{SAKAUE2017105} solved approximately the problem of maximizing a monotone $k$-submodular subject to a  matroid constraint with the approximation ratio of $\frac{1}{2}$ and the query complexity of $O(rn(IO + k\cdot EO))$ based on a greedy technique, 
where $r$ and $IO$ denote the rank and the time of independence oracle of the given matroid, respectively, and $EO$ denotes the time of estimated oracle of the $k$-submodular function.
Rafiey and Yoshida~\cite{Fast-and-Private2020} gave the first asymptotically tight $\frac{1}{2}$-approximation algorithm for this problem that preserves differential privacy, and its complexity $O(kn\ln r\ln \frac{r}{\gamma})$ is almost linear with the function evaluation, where $\gamma$ is the failure probability of this algorithm.
Wang et al.~\cite{Wang2022WeaklyKM} applied a greedy algorithm to maximize a weakly $k$-submodular function, where the constraint is also about a matroid. 
For the problem to maximize a monotone $k$-submodular function under the intersection of a knapsack and $m$ matroid constraints, Yu et al.~\cite{Yu2023} proposed a nested and local search algorithm to obtain an approximation ratio of $\frac{1}{m+2}(1-e^{-(m+2)})$.
And for non-monotone subjective function, they obtained a $\frac{1}{m+3}(1-e^{-(m+3)})$-approximation algorithm.
Sun et al.~\cite{sunyunjing2022} also gave a deterministic algorithm with an approximation factor of $1/3$ for the non-monotone case with the same query complexity as the one of monotone case~\cite{SAKAUE2017105}.
In addition, based on the deterministic algorithm, they designed a randomized algorithm for this problem, which was able to obtain an approximation guarantee of $1/3$ and query complexity of $O(n(IO\log \frac{r}{\e_{1}}+k\cdot EO\log \frac{r}{\e_{2}})\log r)$ with a probability of $1-\e$, where $\e = \max\{\e_{1},\e_{2}\}$.
For more results about $k$-submodular functions, one can see~\cite{pmlr-v157-matsuoka21b,Pham2021StreamingAF,Qian8026144Multi-objective,TANG202286secretary,Yu2022}.

In this paper, we mainly study the fast algorithm for maximizing a $k$-submodular function under a matroid constraint as follows:
\begin{itemize}
\item We design a fast algorithm by using the Threshold-Decreasing method combined with greedy ideas to maximize $k$-submodular functions in both monotone and non-monotone cases under a matroid constraint.
We improve the complexity of $O(rn(IO +k\cdot EO))$ to $O(\frac{n(IO +k\cdot EO)}{\e} \log \frac{r}{\e})$ while obtaining an almost tight approximation guarantee, achieving no linear dependence on the rank $r$ of the given matroid, where $IO$ and $EO$ denote the number of oracles to evaluate whether a subset is an independent set and to compute the function value of $f$, respectively.
\item Based on our result about the matroid constraint, two corollaries about maximizing $k$-submodular functions subject to a total size constraint are obtained, since the constraint of total size constraint can be viewed as a uniform matroid.
\end{itemize}

We organize our paper as follows. We first introduce the useful notation and definitions for $k$-submodular functions and matroid in Section~\ref{Pre}. Then we present the fast algorithm and the proof of the approximation guarantee as well as query complexity about the problem of maximizing $k$-submodular functions under a matroid constraint in Section~\ref{sec_M}. Finally, we summarize our paper in Section \ref{DG}.

\section{Preliminaries}\label{Pre}
In this section, we mainly introduce some notations and properties of $k$-submodular functions.
For any $ \boldsymbol{p} = (P_{1},P_{2},\dots,P_{k}) \in ( k+1 ) ^E $,
we define the \emph{support set} of $\boldsymbol{p}$ as a collection of all elements contained in $P_{i} (\forall i \in [k])$, that is, $supp( \boldsymbol{p}) = \bigcup_{i \in [k]} P_{i}$.
Besides, we use $|supp( \boldsymbol{p} )|$ to denote the size of $ \boldsymbol{p}$.
We also denote $\boldsymbol{0} = (\emptyset,\emptyset,\ldots,\emptyset)$ and assume that $f(\boldsymbol{0}) = 0$.
There is another vector representation of the domain of $k$-submodular function.
For the convenience of analysis, we will identify the sets of $k$-tuples with $n$-vectors $ \{ 0,1,\ldots, k \}^{E} $.
We define $ \boldsymbol{p} \in \{0,1,\ldots, k\}^{E} $ to be an $n$-dimensional vector as follows:
If the element $e$ belongs to some $i^{\textrm{th}}$ set $P_{i}$, we define $\boldsymbol{p}(e) = i$,
otherwise, i.e., $e\notin supp(\boldsymbol{p})$, then write $\boldsymbol{p}(e) = 0$. Given $ \boldsymbol{p} \in \{ 0,1,\ldots,k \}^{E} $, then associate $ (P_{1},P_{2},\ldots,P_{k}) \in (k+1)^{E} $ by:
\[
P_{i} = \{ e \in E | \boldsymbol{p} (e) = i \}, \quad \textrm{for} \,\, \textrm{any} \,\, i \in [k].
\]
We sometimes abuse the notations, and simply write $ \boldsymbol{p} = (P_{1},P_{2},\ldots,P_{k}) $ by regarding a vector $\boldsymbol{p}$ as $k$ disjoint subsets of $E$.
Moreover, we use $\textbf{1}_{e}$ to represent a unit vector whose coordinate is $1$ on $e$ and other coordinates are $0$.

Taking any $k$-tuples $ \boldsymbol{p} = (P_{1},P_{2},\ldots,P_{k})$ and $\boldsymbol{q} = (Q_{1},Q_{2},\ldots,Q_{k})$, denote by $ \boldsymbol{p} \preceq \boldsymbol{q} $ if $ P_{i} \subseteq Q_{i} $ for all $ i \in [k]$.
We say that a $k$-submodular function $f$ is \emph{monotone} if $ f ( \boldsymbol{p} ) \le f ( \boldsymbol{q} )$ for any $\boldsymbol{p} \preceq \boldsymbol{q}$.
Then the \emph{marginal gain} of adding element $e \notin supp(\bp)$ to the $i^{\textrm{th}}$ set $P_{i}$ of $\boldsymbol{p}$ is denoted as
\[
\Delta_{e,i} f ( \boldsymbol{p} ) = f ( P_{1},\ldots,P_{i} \cup \{e\},\ldots,P_{k} ) - f ( P_{1},\ldots,P_{i},\ldots,P_{k} ).
\]
Thus we can define the \emph{orthant submodularity} of a set function $f$ if it satisfies the following condition,
\[
\Delta_{e,i} f ( \boldsymbol{p} ) \ge \Delta_{e,i} f ( \boldsymbol{q} ), \forall \boldsymbol{p},\boldsymbol{q} \in (k+1)^{E} \, \textrm{with} \, \boldsymbol{p} \preceq \boldsymbol{q}, e \notin supp(\bq),\forall i \in [k].
\]
Similarly, we define the \emph{pairwise monotonicity} of $f$ if it satisfies,
\[
\Delta_{e,i} f ( \boldsymbol{p} ) + \Delta_{e,j} f ( \boldsymbol{p} ) \ge 0, \forall \boldsymbol{p} \in (k+1)^{E} \, \textrm{with} \, e \notin supp(\bp), \, \textrm{and} \, i \ne j \in [k].
\]

We present the known results for $k$-submodular functions in the following lemmas. 
\begin{theorem}\label{thm1}
(\cite{WardZivny2016Beyond}).
A real value set function $f$ defined on $(k+1)^{E}$ is $k$-submodular if and only if $f$ not only satisfies orthant submodularity but also satisfies pairwise monotonicity.
\end{theorem}

\begin{lemma}\label{OS}
(\cite{TANG202228}). 
Assume that $f$ is a $k$-submodular function, then for any $k$-tuples $\bp,\bq$ with $\bp \preceq \bq$, we have
\[
f(\bq) - f(\bp) \le \sum_{e \in supp(\bq) \setminus supp(\bp)} \Delta_{e,\bq(e)} f(\bp)
\]
\end{lemma}

In the next part, we will introduce some definition and known results about matroid:
\begin{definition}
(\cite{KV2011}). 
Given a finite ground set $E$, $\mathcal{F} \subseteq 2^{E}$, we call the pair $(E, \mathcal{F})$ a matroid if it satisfies the following conditions. 
\begin{enumerate}[(a)]
  \item\label{a} $\mathcal{F}$ contains the empty set;
  \item\label{b} Assume $B$ belongs to $\mathcal{F}$, then any subset of $B$ also belongs to $\mathcal{F}$.
  \item For any two sets $A, B$ in $\mathcal{F}$ with $|A| < |B|$, then there exists $e \in B \setminus A$ such that $A \cup \{e\}$  also belongs to $\mathcal{F}$.
\end{enumerate}
\end{definition}
Given a matroid $(E, \mathcal{F})$, 
any element $B \in \mathcal{F}$ is called an independent set. Given an independent set $B \in \mathcal{F}$, if there is no bigger independent set containing $B$, then we say $B$ is a \emph{basis} of the matroid, and the cardinality of $B$ is called the \emph{rank} of the matroid, denoted by $r$. 

There is a useful result about the independent sets for exchanging elements.
\begin{lemma}\label{lem3}
(\cite{SAKAUE2017105}). 
Given a matroid $(E, \mathcal{F})$.
Suppose that $ A $ is an independent set and $ B$ is a basis with $ A \subsetneq B $. 
Then, for any element $ e \notin A $ satisfying that $ A \cup \{ e \} $ is an independent set,
there exists a corresponding element $ e' \in B \setminus A $ such that $B \setminus \{ e' \} \cup \{ e \}$ is also a basis.
\end{lemma}

Given a matoid $(E, \mathcal{F})$, we study the problem of maximizing a $k$-submodular function subject to that the support of the $k$-tuple is an independent set.
That is to say, the question we study is:
\be\label{matroid-k}
\max_{\bp \in (k+1)^{E}} f(\bp)
\ee
\[
{\rm s.t.} \quad  supp(\bp) \in \mathcal{F}.
\]

Assuming that the optimal value of this problem is $OPT$,
thus the maximal optimal solution can also be defined as:
\[
\argmax_{\bp} \{|supp(\bp)|: f(\bp) = OPT, supp(\bp) \in \mathcal{F}\}.
\]

It can be shown that the size of the maximal optimal solution of this problem is the same as the rank of the matroid in the following lemmas.

\begin{lemma}\label{lem2} (\cite{SAKAUE2017105}). 
Given a matroid $(E, \mathcal{F})$ with rank $r$ and the $k$-submodular function $f$ is monotone, then the size of problem~(\ref{matroid-k}) is $r$.
\end{lemma}

\begin{lemma}\label{lem4}
(\cite{sunyunjing2022}). 
Given a matroid $(E, \mathcal{F})$ with rank $r$ and the $k$-submodular function $f$ is monotone, then the size of problem~(\ref{matroid-k}) is also $r$.
\end{lemma}


\section{Maximizing a $k$-submodular Function under a Matroid Constraint}\label{sec_M}
In this section, we introduce our main result about the problem of maximizing $k$-submodular functions under a matroid constraint. We first present the fast algorithm, then give the analysis of this algorithm, at last we will show that the results about the problems with a total size constraint can be obtained as corollaries.

\subsection{Fast Algorithm for Maximizing a $k$-submodular Function under a Matroid Constraint}
For the problem of maximizing a $k$-submodular function under a matroid constraint, we introduce a fast algorithm by reducing the threshold value constantly in Algorithm~\ref{alg1}.
The final output solution $\bp$ can be obtained in the following way:
We first let $\bp$ be $\boldsymbol{0}$,
then judge each of the elements in the ground set $E$ in a randomized order as to whether they should be added to $supp(\bp)$ or not.
If the current $supp(\bp)$ can remain an independent set with its marginal gain above the current threshold after adding an element $e$, then we add this element to $supp(\bp)$, otherwise put it back into $E$.
If there are no points that meet the threshold to be selected, the threshold will be reduced by a factor of $ 1-\epsilon $.
By continuously lowering the threshold, all points that meet each threshold are selected until reaching a specific threshold $ \frac{\e}{2r}d $ ($d$ is the maximal single-point value of $f$) or selecting enough $r$ elements.
When the threshold is below $ \frac{\e}{2r}d $, no more points will be added even if not enough $r$ elements have been selected.
Next, we will discuss this problem in monotone and non-monotone cases. Moreover,
Our algorithm will reduce the complexity of the algorithm while ensuring that the approximation ratio is about the same as the greedy algorithm, making it sub-linear dependent of $r$.

\begin{algorithm}[ht]

	\renewcommand{\algorithmicrequire}{\textbf{Input:}}
    \renewcommand{\algorithmicensure}{\textbf{Output:}}
	\caption{Decreasing-Threshold procedure under Matroid Constraint}
	\begin{algorithmic}[1]\label{alg1}

		\REQUIRE A $k$-submodular function $ f:(k+1)^E \to \mathbb{R} $, a matroid $(E, \mathcal{F})$ with rank $r$, $ \e \in (0,1) $.
        \ENSURE A solution $ \bp $ satisfied $supp (\bp) \in \mathcal{F} $.
        \STATE $ \bp \gets \boldsymbol{0} $.
        \STATE $ w \gets d = \max _{e \in E, i \in [k]} f(i\textbf{1}_{e}),  F(\bp) \gets E$
        \WHILE {$w > \frac{(1-\e)\e}{2r}d$ and $F(\bp) \ne \emptyset$}    \label{WHILE}          
               \FOR {$e \in F(\bp)$}       \label{FOR}     
                    \STATE $ i \gets \argmax_{i \in [k]} \Delta_{e,i} f(\bp) $     \label{EO}      
                    \IF {$ \Delta_{e,i} f(\bp) \ge w $}
                        \STATE $ \bp ( e ) \gets i $
                        \STATE construct a collection of feasible elements $F(\bp)=\{e\in E| supp(\bp)\cup\{e\}\in \mathcal{F}\}$ using independent oracle  \label{IO}    
                    \ENDIF
                \ENDFOR
                    \STATE $w=w(1-\e)$
         \ENDWHILE
        \RETURN $ \bp $
	\end{algorithmic}
\end{algorithm}

\subsection{Proof of the approximation guarantee as well as query complexity}

Let $\boldsymbol{o}$ be a maximal optimal solution and $\bp$ be the returned solution by Algorithm~\ref{alg1}. Assume that after continuous iteration to select points, the algorithm selects $t ( t \le r)$ elements in total, then we gradually establish the sequence  $\boldsymbol{p}^{0}, \boldsymbol{p}^{1}, \ldots ,\boldsymbol{p}^{t}$ in the following way.
Assume that $\bp^{0}=\textbf{0}$ and suppose that the $j$-th selected point and its position is
$( e_{j},i_{j} )$, then we define $\bp^{j}=\bp^{j-1}+i_j\textbf{1}_{e_j}, j\in [t]$. It is obvious to know that $\bp=\bp^t$.

To analyze the relationship between the maximal optimal solution $ \boldsymbol{o} $ and the output solution $ \boldsymbol{p} $ of the algorithm, we let $\boldsymbol{o}^{0}=\boldsymbol{o}$ and construct the corresponding sequence of $\boldsymbol{o}^{0}, \boldsymbol{o}^{1}, \ldots ,\boldsymbol{o}^{t}$ based on the sequence of $ \boldsymbol{p}^{0}, \boldsymbol{p}^{1}, \ldots ,\boldsymbol{p}^{t} $.
By modifying only one element in $\bo$ at a time, we keep at least $M-1$ elements between two adjacent vectors of $\boldsymbol{o}^{0}, \boldsymbol{o}^{1}, \ldots ,\boldsymbol{o}^{t}$ identical (including elements and positions).
This ensures that the elements $e_{1}, \dots, e_{t}$ selected by the algorithm end up in the same position in $\bo^{t}$ as in $\bp$.
Finally, the relation between the output solution $\bp$ and the maximal optimal solution $\bo$ is established.
Through the following construction procedure we can ensure that $\boldsymbol{o}^{0}, \boldsymbol{o}^{1}, \ldots ,\boldsymbol{o}^{t}$ are all independent sets.

According to the previous assumption, $e_{j}$ is the $j$-th  element selected by the algorithm for each $j \in [t]$.
The process of constructing $\bo^{j}$ by $\bo^{j-1}$ is divided into two steps:
First, select an element in $supp(\bo^{j-1})$ and take it out, then put the element $e_{j}$ chosen by Algorithm~\ref{alg1} into the $i_{j}$-th set of $\bo^{j-1}$.
We mark the elements removed from $supp(\bo^{j-1})$ as $o_{j}$.
Obviously, the choice of $o_{j}$ depends on whether element $e_{j}$ is in the support set of the optimal solution $\bo$.
If $e_{j}\in supp(\bo^{j-1})$, take $o_{j} = e_{j}$.
Otherwise according to Lemma \ref{lem3}, there exists some $o$ in $supp(\bo) \setminus supp(\bp^{j-1})$ such that $supp(\bo) \setminus \{o\} \cup \{e_{j}\} \in \mathcal{F}$, then we denote this element $o$ as $o_j$.
For the sake of discussion, we define a new notation $\bo^{j-\frac{1}{2}}$ by removing $o_{j}$ from the support of $\bo^{j-1}$, i.e.,  \[
\bo^{j-\frac{1}{2}}: = \bo^{j-1} - \bo^{j-1}(o_{j}) \textbf{1}_{o_{j}}.
\]
Then $\bo^{j}$ can be constructed by adding $(e^j,i^j)$ to  $\boldsymbol{o}^{j-\frac{1}{2}}$, i.e.,
\[
\bo^{j}:= \bo^{j-\frac{1}{2}} + i_{j} \textbf{1}_{e_{j}}.
\]
It is not difficult to obtain $ \boldsymbol{p} = \boldsymbol{p}^{t} \preceq \boldsymbol{o}^{t} $
and $ \boldsymbol{p}^{j-1} \preceq \boldsymbol{o}^{j-\frac{1}{2}}$.  Moreover, if $t = r$, then $\bp^{r} = \bo^{r}$.

Suppose now that we have selected $ ( e_{j-1},i_{j-1} ) $ and the next element to be chosen is $ ( e_{j},i_{j} ) $.
Assume that the current threshold is $ w $, so we can easily see that,
\begin{align} \label{eq1}
\Delta_{e_{j},i_{j}} f ( \boldsymbol{p}^{j-1} ) \ge w.
\end{align}

Assume that the position of the element $ o_{j} $ in the maximal optimal solution $ \boldsymbol{o} $ is $ i^{*} $,
i.e.,
$ \boldsymbol{o}^{j-1}(o_{j}) = i^{*} $, and the last element chosen under the previous threshold $\frac{w}{1-\e}$ is $ e_{j'} $.
The reason that $ {o_{j}} $ was not selected under that threshold is that the gain added $ {o_{j}} $ to $ \bp^{j'} $ does not reach $\frac{w}{1-\e}$.
At the same time, $ \boldsymbol{p}^{j'} \preceq \boldsymbol{p}^{j-1} $,
so we have,
\begin{equation} \label{eq2}
\Delta_{o_{j},i^*} f ( \boldsymbol{p}^{j-1} ) \le \Delta_{o_{j},i^*} f ( \boldsymbol{p}^{j'} ) < \frac{w}{1-\epsilon}.
\end{equation}

\begin{claim} \label{cla3}
If the objective function $f$ is monotone, we have
\[
f( \boldsymbol{o}^{j-1} ) - f( \boldsymbol{o}^{j} ) < \frac{1}{1-\e} \big( f ( \boldsymbol{p}^{j} ) - f ( \boldsymbol{p}^{j-1} ) \big),\forall j \in [t].
\]
\end{claim}

\begin{prf}
Referring to the construction process of the sequence $ \boldsymbol{o}^{0}, \boldsymbol{o}^{1}, \ldots ,\boldsymbol{o}^{t} $ in Algorithm \ref{alg1}, we can see that for each $j \in [t] $, there is $ \boldsymbol{o}^{j-\frac{1}{2}} \preceq \boldsymbol{o}^{j} $, and by the monotonicity, we have,
\begin{equation} \label{eq3}
f( \boldsymbol{o}^{j-\frac{1}{2}} ) - f ( \boldsymbol{o}^{j} ) \le 0.
\end{equation}
Combining (\ref{eq1}) to (\ref{eq3}) we have,
\begin{align*}
f ( \boldsymbol{o}^{j-1} ) - f ( \boldsymbol{o}^{j} )
&= f ( \boldsymbol{o}^{j-1} ) - f ( \boldsymbol{o}^{j-\frac{1}{2}} ) + f ( \boldsymbol{o}^{j-\frac{1}{2}} ) - f ( \boldsymbol{o}^{j} )   \\
&\le f ( \boldsymbol{o}^{j-1} ) - f ( \boldsymbol{o}^{j-\frac{1}{2}} )  &(\textrm{monotonicity}) \\
&= \Delta_{o_{j},i^*} f ( \boldsymbol{o}^{j-\frac{1}{2}} )    \\
&\le \Delta_{o_{j},i^*} f ( \boldsymbol{p}^{j-1} )     &(\textrm{submodularity})   \\
&< \frac{w}{1-\e} &(\textrm{due\, to\,(\ref{eq2})}) \notag  \\
&\le \frac{1}{1-\e} \Delta_{e_{j},i_{j}} f ( \boldsymbol{p}^{j-1} ) &(\textrm{due\,to\,(\ref{eq1})})   \\
&= \frac{1}{1-\e} ( f ( \boldsymbol{p}^{j} ) - f ( \boldsymbol{p}^{j-1} ) ).
\end{align*}
\end{prf}

\begin{claim} \label{cla1}
If the objective function $f$ is non-monotone,, we have
\[
f(\boldsymbol{o}^{j-1} ) - f( \boldsymbol{o}^{j} ) < \frac{2-\e}{1-\e}(f(\boldsymbol{p}^{j}) - f(\boldsymbol{p}^{j-1})), \forall j \in [t].
\]
\end{claim}

\begin{prf}
The idea of the proof of this claim is roughly the same as that of Claim \ref{cla3}.
The difference is that we resort to the element and position pair $ ( e_{j},i' ) $ to complete the transition, where $ i' $ denotes any position that differs from $ i_{j} $.
Since $ \boldsymbol{o}^{j} = \boldsymbol{o}^{j-\frac{1}{2}} + i_{j} \textbf{1}_{e_{j}} $, due to the pairwise monotonicity, there is:
\begin{equation} \label{eq4}
\Delta_{e_{j},i'} f ( \boldsymbol{o}^{j-\frac{1}{2}} ) + \Delta_{e_{j},i_{j}} f ( \boldsymbol{o}^{j-\frac{1}{2}} ) \ge 0.
\end{equation}
Since the position of element $ e_{j} $ is greedily selected, therefore,
\begin{equation} \label{eq5}
\Delta_{e_{j},i'} f ( \boldsymbol{p}^{j-1} ) \le  \Delta_{e_{j},i_{j}} f ( \boldsymbol{p}^{j-1} ).
\end{equation}
Thus combining (\ref{eq1}), (\ref{eq2}), (\ref{eq4}) and (\ref{eq5}) we have:
\begin{align*}
&f ( \boldsymbol{o}^{j-1} ) - f ( \boldsymbol{o}^{j} )\\
&= f(\bo^{j-\frac{1}{2}} + i^* \textbf{1}_{o_{j}})  - f(\bo^{j-\frac{1}{2}} + i_{j} \textbf{1}_{e_{j}})    \\
&= f(\bo^{j-\frac{1}{2}} + i^* \textbf{1}_{o_{j}}) - f(\bo^{j-\frac{1}{2}}) + f(\bo^{j-\frac{1}{2}} + i' \textbf{1}_{e_{j}}) - f(\bo^{j-\frac{1}{2}}) \\
&\quad- f(\boldsymbol{o}^{j-\frac{1}{2}} + i' \textbf{1}_{e_{j}}) + f(\boldsymbol{o}^{j-\frac{1}{2}}) - f(\bo^{j-\frac{1}{2}} + i_{j} \textbf{1}_{e_{j}}) + f(\bo^{j-\frac{1}{2}})   \\
&= \Delta_{o_{j},i^*} f( \boldsymbol{o}^{j-\frac{1}{2}} ) + \Delta_{e_{j},i'} f( \boldsymbol{o}^{j-\frac{1}{2}} ) - \Delta_{e_{j},i'} f( \boldsymbol{o}^{j-\frac{1}{2}} ) - \Delta_{e_{j},i_{j}} f( \boldsymbol{o}^{j-\frac{1}{2}} ) \\
&\le \Delta_{o_{j},i^*} f( \boldsymbol{o}^{j-\frac{1}{2}} ) + \Delta_{e_{j},i'} f( \boldsymbol{o}^{j-\frac{1}{2}} ) \quad\quad (\textrm{due to\,(\ref{eq4})})  \\
&\le \Delta_{o_{j},i^*} f( \boldsymbol{p}^{j-1} ) + \Delta_{e_{j},i'} f( \boldsymbol{p}^{j-1} )  \quad\quad (\textrm{submodularity})\\
&< \frac{w}{1-\e} + \Delta_{e_{j},i'} f( \boldsymbol{p}^{j-1} ) \quad\quad \quad\quad \quad\quad (\textrm{due\,to\,(\ref{eq2})}) \\
&\le \frac{1}{1-\e} \Delta_{e_{j},i_{j}} f( \boldsymbol{p}^{j-1} ) + \Delta_{e_{j},i_{j}} f( \boldsymbol{p}^{j-1} )  \quad\quad (\textrm{due \,to\,(\ref{eq1})\,and\,(\ref{eq5})})  \\
&= \frac{2-\e}{1-\e} ( f ( \boldsymbol{p}^{j} ) - f ( \boldsymbol{p}^{j-1} ) ).
\end{align*}
\end{prf}

\begin{theorem} \label{thm2}
For the monotone $k$-submodular  maximization problem under total size constraint, Algorithm \ref{alg1} yields a $(\frac{1}{2}-\e)$-approximation solution.
And for the non-monotone case, the approximation of the returned solution from Algorithm \ref{alg1} is $ \frac{1}{3}-\e$.
In both cases, the query complexity of $f$ and independent sets is  $O(\frac{n(k\cdot EO + IO)}{\epsilon} \log \frac{r}{\epsilon})$.
\end{theorem}

\begin{prf}
Here, we only prove the monotone case, and similar analysis method can be applied to the non-monotone case.

From Claim \ref{cla3}, $ \forall j \in [t] $,
\[
f ( \boldsymbol{o}^{j-1} ) - f( \boldsymbol{o}^{j} )
<
\frac{1}{1-\e} ( f ( \boldsymbol{p}^{j} ) - f ( \boldsymbol{p}^{j-1} ) ).
\]
Thus,
\begin{align*}
\sum_{j=1}^{t} ( f ( \boldsymbol{o}^{j-1} ) -f ( \boldsymbol{o}^{j} ) )
<
\sum_{j=1}^{t} \frac{1}{1-\e} ( f ( \boldsymbol{p}^{j} ) - f ( \boldsymbol{p}^{j-1} ) ),
\end{align*}
which implies,
\begin{align}\label{eq6}
f ( \boldsymbol{o} ) - f ( \boldsymbol{o}^{t} )
&< \frac{1}{1-\e} ( f ( \boldsymbol{p}^{t} ) - f ( \boldsymbol{p}^{0} ) )   \notag \\
&= \frac{1}{1-\e} ( f ( \boldsymbol{p}^{t} ) - f ( \boldsymbol{0} ) )
= \frac{1}{1-\e} f ( \boldsymbol{p}^{t} ).
\end{align}
According to the termination condition of the algorithm \ref{alg1}, we will analyze the approximation ratio in following two cases.

\textbf{Case 1}: If $ t < r $.

In this case, the algorithm stops when only $t$ elements are selected.
At this time, the marginal gain of any element in $ E \setminus supp(\boldsymbol{p}^{t})$ added to $\boldsymbol{p}^{t}$ is less than $\frac{\e d}{2r}$.
Take any $ e \in supp(\bo) \setminus supp(\bp^{t})$,\\
\begin{align}\label{eq10}
\Delta_{e,\boldsymbol{o}(e)}f(\boldsymbol{p}^{t}) \le \frac{\e d}{2r}.
\end{align}
By (\ref{eq6}), Lemma \ref{OS} and (\ref{eq10}) we have,
\begin{align*}
f ( \boldsymbol{o} ) &< f ( \boldsymbol{o}^{t} ) + \frac{1}{1-\e} f ( \boldsymbol{p}^{t} )  &(\textrm{due\,to\,(\ref{eq6})})  \\
&= f ( \bo^{t} ) - f ( \bp^{t} ) + \frac{2-\e}{1-\e} f ( \bp^{t} )     \\
&\le \sum_{ e \in supp (\bo^{t}) \setminus supp (\bp^{t}) } \Delta_{ e,\bo(e) } f( \bp^{t} )  + \frac{2-\e}{1-\e} f ( \bp^{t} ) &(\textrm{due\,to\,Lemma\,\ref{OS}})\\
&\le M \frac{\e d}{2M} + \frac{2-\e}{1-\e} f ( \boldsymbol{p}^{t} ) &(\textrm{due\,to\,(\ref{eq10})}) \\
&\le \frac{\e}{2} f ( \boldsymbol{o} ) + \frac{2-\e}{1-\e} f ( \boldsymbol{p}^{t} ).
\end{align*}
Due to the construction process of sequence $ \boldsymbol{p}^{0}, \boldsymbol{p}^{1}, \ldots ,\boldsymbol{p}^{t} $, we can know that $ \boldsymbol{p}^{t} = \boldsymbol{p} $.
Thus,
\[
( 1-\frac{\e}{2} ) f ( \boldsymbol{o} ) \le \frac{2-\e}{1-\e} f ( \boldsymbol{p} ).
\]
Then we obtain,
\[
f ( \boldsymbol{p} ) \ge \frac{1-\e}{2} f ( \boldsymbol{o} ) \ge (\frac{1}{2}-\e) f ( \boldsymbol{o} ).
\]
\textbf{Case 2}: If $t = r$.

In this case, we have $ \boldsymbol{o}^{r} = \boldsymbol{p}^{r}  = \boldsymbol{p}$.
According to (\ref{eq6}),
\[
f(\boldsymbol{o}) - f(\boldsymbol{o}^{r}) = f(\boldsymbol{o}) - f(\boldsymbol{p}^{r}) < \frac{1}{1-\e} f(\boldsymbol{p}^{r}).
\]
This implies,
\[
f(\boldsymbol{p}) \ge \frac{1-\e}{2-\e} f(\boldsymbol{o}) \ge (\frac{1}{2}-\e) f ( \boldsymbol{o} ).
\]
Then we get the approximation ratio of $\frac{1}{2}-\e$.

It is easy to see that the number of iterations in the outer loop in Line \ref{WHILE} of Algorithm~\ref{alg1} is $O( \frac{1}{\e} \log \frac{r}{\e} )$.
And the number of iterations in the inner \textbf{for} loop in Line \ref{FOR} is $n$.
Suppose that the complexity of the oracle to calculating the value of $f$ in Line \ref{EO} is denoted by $EO$,
and the complexity of the oracle to determine whether a set is an independent set in Line \ref{IO} is denoted by $IO$.
So the running time is,
\[
\frac{1}{\epsilon} \log \frac{r}{\epsilon} \times n \times (k \times EO + IO) = \frac{n(k\cdot EO + IO)}{\epsilon} \log \frac{r}{\epsilon}.
\]
\end{prf}

\subsection{Two corollaries for maximizing $k$-submodular functions under a total size constraints}

Given an upper bound number $B$, the maximization of $k$-submodular functions subject to a total size constraint can be stated as:
\[
\max_{\boldsymbol{p} \in (k+1)^{E}} f(\boldsymbol{p})
\]
\[
{\rm s.t.} \quad |supp ( \boldsymbol{p} ) | \le B.
\]
We know this constraint can be viewed as a uniform matroid by taking
\[
\mathcal{F} = \{ P \subseteq E: |P| \le B\}.
\]

Then we have the following two corollaries:

\begin{corollary}(\cite{Nie2023FAST_size}).
For the monotone $k$-submodular maximization under total size constraint, we can obtain a $(\frac{1}{2}-\e)$-approximation solution with $O( \frac{nk}{\epsilon} \log \frac{B}{\epsilon})$ query complexity.
\end{corollary}

\begin{corollary}
For the non-monotone $k$-submodular maximization under total size constraint, we can obtain a $ (\frac{1}{3}-\e) $-approximation solution with $O( \frac{nk}{\epsilon} \log \frac{B}{\epsilon})$ query complexity.
\end{corollary}

\section{Conclusion}\label{DG}
In this paper, we present a fast algorithm to solve the problem of maximizing monotone and non-monotone $k$-submodular functions under a matroid constraint.
Compared to the greedy algorithm, the advantage of this fast algorithm is that its complexity does not linearly depend on the rank of the given matroid. Moreover, these results can generalize the known result for maximizing monotone $k$-submodular functions subject to a total size constraint.
\section*{Acknowledgements}
The second and fourth authors are supported by Natural Science Foundation of Shandong Province of China (Nos. ZR2020MA029, ZR2021MA100). The third author is supported by National Science Foundation of China (No. 12001335).





\end{document}